# Atomic-scale effect of 2D π-conjugated metal-organic frameworks as electrocatalysts for $CO_2$ reduction reaction towards highly selective products

*Ran Wang [ab], Chaozheng He [b*], Weixing Chen [b*], Qingquan Kong [a*], Thomas Frauenheim [ac]*

[a] *Institute for Advanced Study, Chengdu University, Chengdu 610106, P. R. China*

[b] *Institute of Environmental and Energy Catalysis, Shaanxi Key Laboratory of Optoelectronic Functional Materials and Devices, School of Materials Science and Chemical Engineering, Xi'an Technological University, Xi'an 710021, P. R. China*

[c] *School of Science, Constructor University, Bremen 28759, Germany*

*Corresponding authors.

*E-mail addresses: hecz2019@xatu.edu.cn (C. He); chenwx@xatu.edu.cn (W. Chen); kongqingquan@cdu.edu.cn (Q. Kong);*

**Abstract**

Electrocatalytic $CO_2$ reduction technology is key to mitigating greenhouse gas emissions and the energy crisis. However, controlling the selectivity of $CO_2RR$ products at low overpotential remains a challenge. In this paper, we predicted five high-performance $CO_2RR$ electrocatalysts with different product-specific selectivities at the theoretical level based on the advantages of the compositional structure and the tunable pore size of 2D π-conjugated MOFs. In addition, through the reaction mechanism and electronic structure analysis, we found that the synergistic interaction between metal atoms and organic linkers of 2D MOFs can effectively regulate the electronic structure of the active center. Their pore size as well as the diversity of carbon materials can regulate the spin magnetic moments of the metal atoms, thus affecting the improvement of their catalytic performance. Meanwhile, the oxygen or carbon affinity of the catalyst surface determines the differences in the formation of key intermediates, which ultimately determines the reaction path and product selectivity. These insights we present will be useful for the development and design of highly active $CO_2RR$ electrocatalysts.

# 1 Introduction

Large emissions of $CO_2$ gas exacerbate environmental problems, such as the greenhouse effect and ocean acidification.[1] Electrocatalytic $CO_2$ reduction reaction ($CO_2$RR) can drive the production of high value-added chemicals and fuels, which can effectively alleviate energy stress and environmental degradation by using these renewable energies.[2-4] However, the actual yields of final products from $CO_2$RR are not satisfactory, because of the stable C=O bonds in linear non-polar $CO_2$ molecule resulting in the slow reaction kinetics, the low Faraday efficiency due to competing the hydrogen evolution reaction (HER), and many by-product outputs.[5-7] Therefore, finding and designing an efficient $CO_2$RR electrocatalyst with high activity and specific selectivity remains a challenge.

The carbon-based single-atom catalysts (SACs) have promising applications in a variety of electrochemical reactions due to their high electrical conductivity, large atomic utilization and low costs.[8-10] The performance of active metal centers can also be controlled by adjusting the local coordination environment, such as changing the coordination atoms and the number of coordination sites and inducing defects.[11, 12] He *et al.* reported that an N-doped carbon-based Pd SACs could reduce $CO_2$ to CO, in which Pd-$N_4$ as an active center was more favorable than that of a single Pd.[13] Yang *et al.* found that the difference in coordination number formed by Cu embedded in $C_3N_4$ could significantly modulate the selectivity of $CO_2$RR, in which the Cu-$N_4$ active center effectively improved the selectivity of $CH_3OH$ by up to 95.5%.[14] In addition, Sun *et al.* investigated a series of TM-$X_4$ (X=S and N) as active centers in carbon-based SACs for $CO_2$RR, where $CoS_4$ could selectively reduce $CO_2$ molecules to form HCOOH molecules, with a low limiting potential of -0.07 V.[15] Although the TM-$X_4$ as active centers in carbon-based SACs is efficient for $CO_2$RR, the role of TM, X, and C configurations in determining $CO_2$RR activity and product selectivity remains elusive and unclear. Moreover, achieving large-scale accurate preparation of TM-$X_4$ SACs remains a major challenge due to the lack of precise and controllable synthetic methods and characterization tools.

Two-dimensional (2D) metal-organic frameworks (MOFs) are composed of metal ions and organic ligands and have well-defined structural features[16]. 2D MOFs have been extensively studied in catalysis due to their well-defined structures, high thermal stability, and fascinating physical and chemical properties.[17] In particular, 2D conjugated MOFs have a scalable π-d conjugated structure, giving them high electrical conductivity, and tunable composition, morphology, and porosity also provide a simple means to manipulate their catalytic properties.[18] Therefore, the specific active centers and coordination environments in 2D conjugated MOFs could provide a viable research platform for understanding the catalytic mechanisms of $CO_2RR$.[19] Liu *et al.* designed and synthesized a highly symmetric hexaazatrinaphthylene (HATNA)-Cu-based conjugated MOFs with $Cu-O_4$ active centers, which exhibits the high selectivity of 78% for the electrocatalytic reduction of $CO_2$ to form $CH_4$.[20] Moreover, Majidi *et al.* reported that the $Cu-O_4$ active center in 2D conjugated MOFs of copper tetrahydroxyquinone could achieve a Faraday efficiency of 91% for the reduction of $CO_2$ to form CO.[21] Zhong et al. found that the $Zn-O_4$ active center formed by Zn atom-doped phthalocyanine-$Cu-O_8$ could effectively improve the selectivity of $CO_2$ reduction to CO.[22] Although these works demonstrate that 2D conjugated MOFs could be used as the electrocatalytic candidates for $CO_2RR$, their atomic-scale effect of the active center in determining catalytic activity and product selectivity remains unclear. Therefore, understanding the key factors that influence the catalytic efficiency of 2D conjugated MOFs will greatly facilitate screening and designing electrocatalysts for $CO_2RR$.

We herein investigated the $CO_2RR$ performance of $TM-X_4$ (TM = Ti, V, Cr, Mn, Fe, Co, Ni and Cu; X =N, O, S and Se) active centers in three types of 2D conjugated MOFs with different porous sizes by performing density functional theory (DFT) calculations. Some 2D conjugated MOFs were confirmed as high-performance electrocatalysts with specific selectivity for $CO_2RR$ by tuning metals, ligands and porous sizes. Meanwhile, the structure–activity relationships of 2D conjugated MOFs for $CO_2RR$ were illustrated by analyzing their electronic properties. Our results provide theoretical insights for the development of 2D conjugated MOFs as efficient electrocatalysts for $CO_2RR$.

**2 Results and discussion**

**2.1 Geometric stability**

Based on the experimentally prepared 2D MOFs[17, 23-25], we investigated three types of 2D MOFs with the $TM-X_4$ active centers but different pore sizes, as shown in Figure 1. Metal ions and phenyl-like molecules (including HIB, HHB, HTB, and HSB) can couple to form two types of 2D MOFs with different pore configurations, namely TM$_3$HXB and TM$_3$HXB$_2$ (X = I, H, T, and S), respectively. And the triphenyl organic ligands (including HITP, HHTP, HTTP, HSTP) can couple metal ions to form the larger pore configurations of 2D MOFs, TM$_3$HXTP$_2$. Although these 2D MOFs have the homogeneous distribution of TM-X$_4$ active centers, the densities of TM-X$_4$ active centers are different due to different organic ligands bonding TM to form pore sizes. Eight 3d metals, including Ti, V, Cr, Mn, Fe, Co, Ni and Cu atoms, were chosen as the targets, based on recent experimental studies.[26-29] Therefore, we would study the CO$_2$RR performance of 96 2D MOFs from eight metals combined with eight organic ligands. The lattice parameters of these 2D MOFs were optimized, as shown in Table S1.

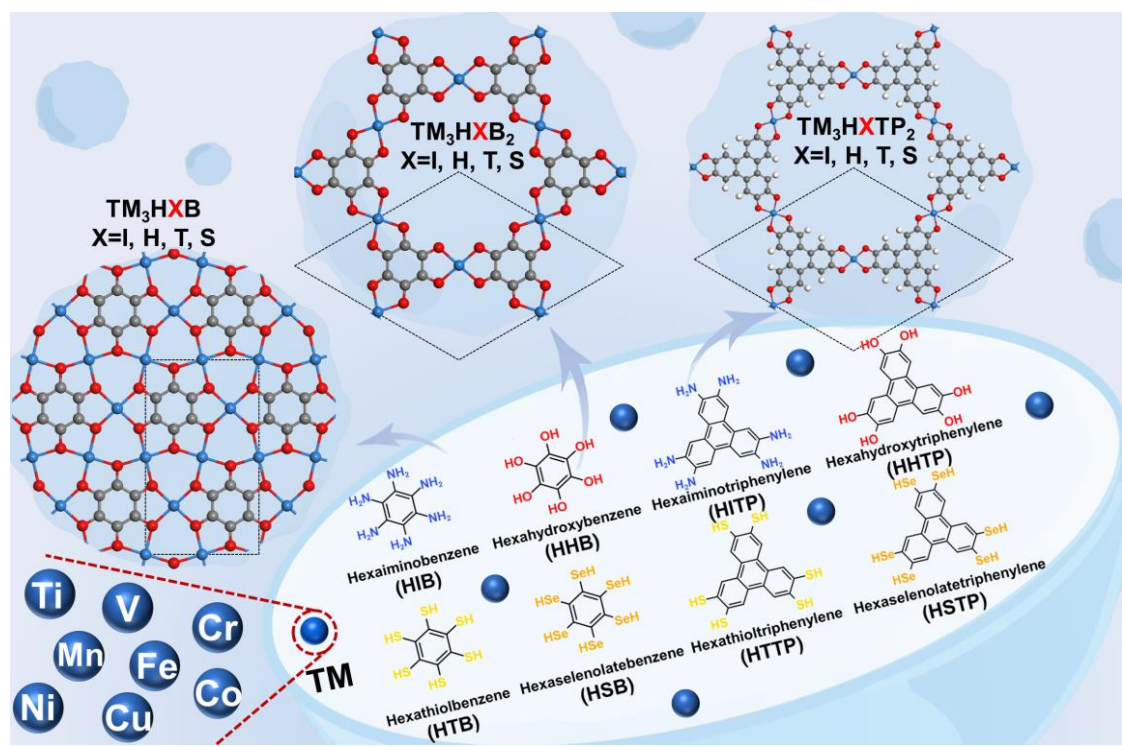

**Figure 1** Schematic structures of metal ions coupled to different organic ligands to form three types of 2D MOFs with the TM-X$_4$ active centres. The blue, red, grey and white spheres represent transition metal (TM), coordinated atom (X), carbon atom and hydrogen atom respectively. The unit cell is denoted by black dotted line.

The stability of the catalyst is the first important factor in catalytic reactions. 2D MOFs are easily susceptible to metal agglomeration during their preparation due to the high surface energy of the metal atoms. Therefore, the structural stability of these 2D MOFs was first assessed by calculating the formation energies of metal atoms and ligands, as shown in Figure

2a. The results show negative formation energies for all structures, indicating that the coupling of metal atoms and ligands is more thermodynamically favorable compared to the formation of metal clusters. In addition, as an electrocatalyst not only thermodynamic stability but also electrochemical stability is required. Therefore, the electrochemical stability of these 2D MOFs have to be considered. The negative dissociation potential of the metals indicates that the metal atoms tend to precipitation and then lead to structural instability under electrochemical conditions. The following studies of the $CO_2RR$ electrochemical process only considered those thermodynamically and electrochemically stable 2D MOFs.

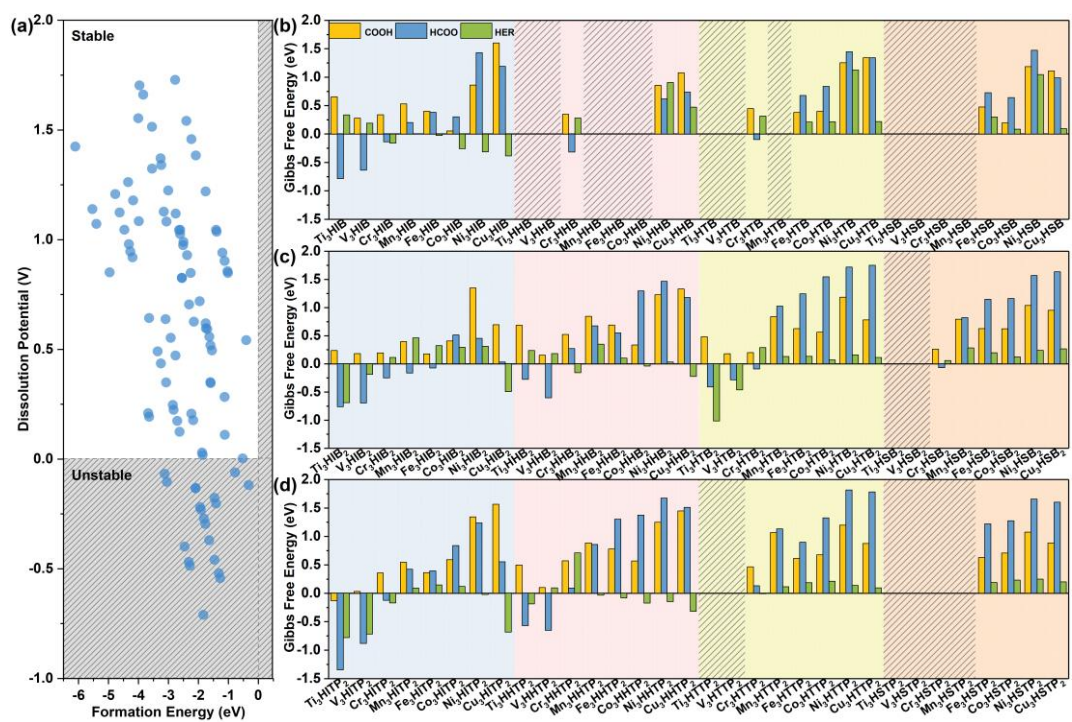

**Figure 2** (a) The formation energies of 2D MOFs and the negative dissociation potential of the metals. Gibbs free energy of the first hydrogenation of $CO_2$ to form COOH and HCOO and HER for (b) $TM_3HXB$, (c) $TM_3HXB_2$, (d) $TM_3HXTP_2$.

## 2.2 Electrocatalytic $CO_2RR$ mechanisms

The reaction mechanisms of $CO_2$ reduction involve are complicated and their products are diverse, due to multiple electron processes and various intermediates. Therefore, the investigation of $CO_2RR$ mechanisms on 2D MOFs is important for further controlling and improving the catalytic activity of 2D MOFs. Because the $CO_2$ molecule is non-polar and has strong thermodynamic stability, the first hydrogenation process in $CO_2RR$ is regarded as the key step, which takes place in the electrolyte solution in the proton-coupled electron transfer (PCET) process, i.e., * + $CO_2$(g) + $H^+$ + $e^-$ → *COOH/*HCOO.[30-32] From the view of the ligands in 2D MOFs, the NH coordination in these 2D MOFs is more favorable for the first hydrogenation of $CO_2$ than that of the O, S, and Se coordination. From the view of the metals in 2D MOFs, the first hydrogenation of $CO_2$ is more difficult with the increased valence electrons in the metal outer layer, as shown in Figure 2b-2d. Moreover, for the competing HER, $TM_3HXB_2$ and $TM_3HXTP_2$ are more favorable for HER than that of $TM_3HXB$, due to the unsaturated coordination environment of ligands in $TM_3HXB_2$ and $TM_3HXTP_2$. Therefore, the catalytic activity of these 2D MOFs could be controlled by metallic atoms, ligands, and their coordination environment.

The proton-electron pair ($H^+$ + $e$) is involved in each elementary step of the $CO_2RR$. The different hydrogenation pathways will lead to various products. Moreover, the competitive selectivity exists between HER and each hydrogenation of $CO_2RR$ under the same reaction conditions, these factors will affect the Faraday efficiency of $CO_2RR$. We first studied HER and the first three hydrogenation processes of $CO_2RR$ on 76 stable 2D MOFs as $CO_2RR$ electrocatalysts with specific selectivity and high performance, as shown in Figure 3.

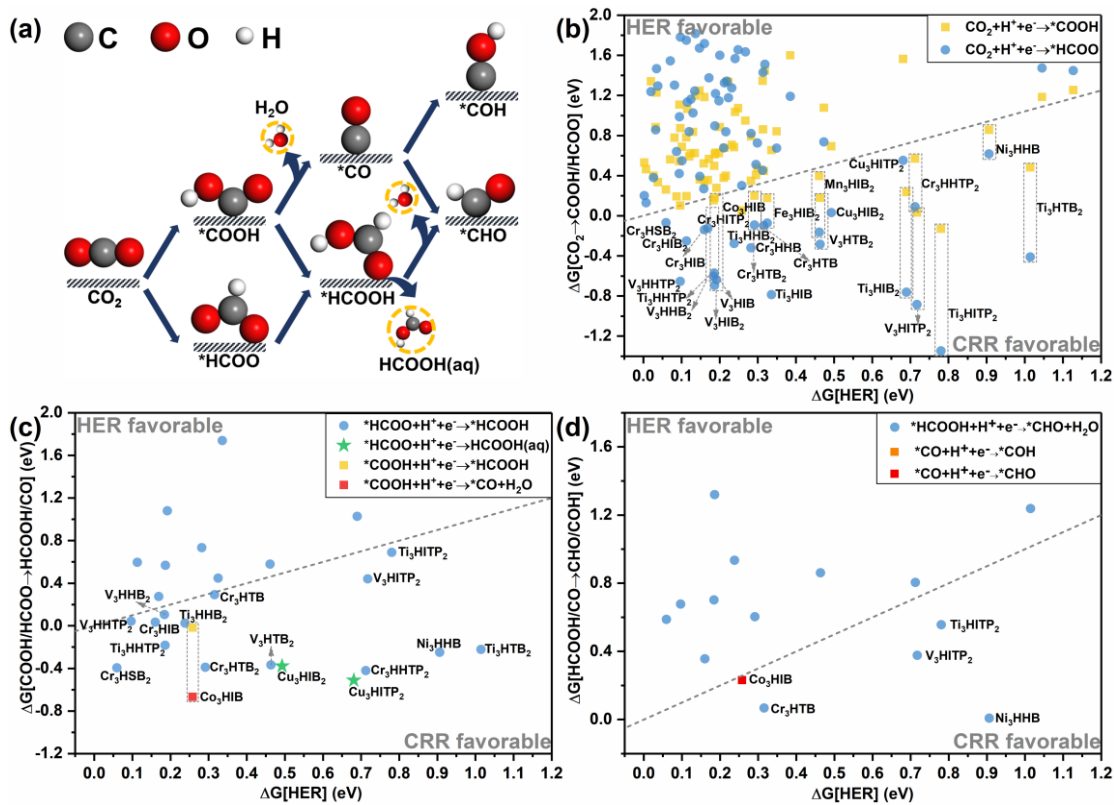

**Figure 3** (a) Schematic diagram of the first three hydrogenation processes of $CO_2$. The comparision of Gibbs free energy between HER and (b) the first hydrogenation, (c) the second hydrogenation, and (d) the third hydrogenation for $CO_2RR$.

The lower Gibbs free energy is more selective, based on the Brønsted–Evans–Polanyi relationship.[33, 34] We screened 26 2D MOFs by comparing the first hydrogenation of $CO_2$ and HER, as shown in Figure 3b. The formation of HCOO on 25 2D MOFs are favorable, except for $Co_3HIB$, which was favorable to form COOH. We then screened 17 2D MOFs by comparing the hydrogenation of HCOO and COOH, as well as HER, as shown in Figure 3c. The H atom combines with either the O atom of HCOO or the C atom of COOH to form HCOOH, while the H atom combines with OH of COOH to form CO and $H_2O$. The strong adsorption of HCOOH and CO can be as intermediates on 2D MOFs for the next hydrogenation processes. In particular, the hydrogenation of HCOO on $Cu_3HIB_2$ and $Cu_3HITP_2$ forms HCOOH, which is directly released and there are no by-products, as shown in Figure 4a and 4b. Meanwhile, the hydrogenation of COOH on $Co_3HIB$ favorably forms CO. Previous works demonstrated *CO is a key intermediate for the formation of C-C coupling into multi-carbon products,[35] which is difficult for most SACs. Therefore, we only considered single carbon products in this work. We finally screened five 2D MOFs, including Ni$_3$HHB, Co$_3$HIB, Cr$_3$HTB, Ti$_3$HITP$_2$, and V$_3$HITP$_2$, by comparing the hydrogenation of HCOOH and CO and HER, as shown in Figure 3d. This hydrogenation process on these 5 2D MOFs favorably forms CHO.

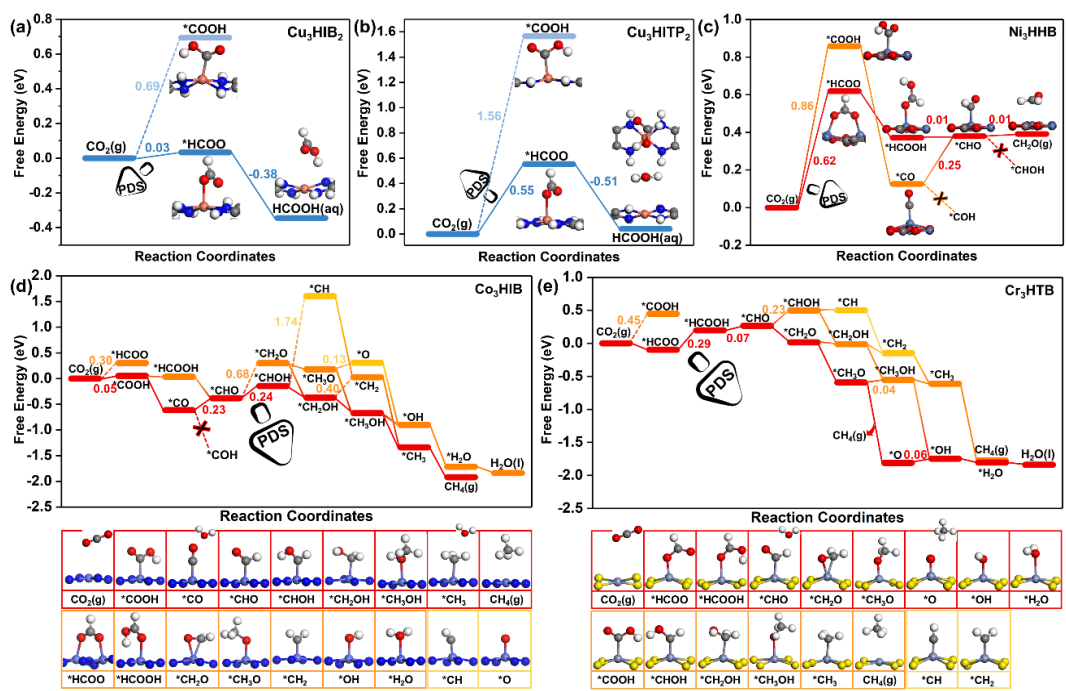

**Firgure 4** The change of Gibbs free energy of CO$_2$RR mechanisms on (a) Cu$_3$(HIB)$_2$, (b) Cu$_3$(HITP)$_2$, (c) Ni$_3$HHB, (d) Co$_3$HIB, (e) Cr$_3$HTB. * represents the intermediate of chemical adsorption.

The CO$_2$RR mechanisms on Ni$_3$HHB, Co$_3$HIB, Cr$_3$HTB, Ti$_3$HITP$_2$ and V$_3$HITP$_2$ were further investigated, as shown in Figure 4c-4e and Figure S1. The H atoms favorably bond to the C atom of the intermediate than that of the O atom on Ni$_3$HHB. In the formation of CH$_2$O by CHO hydrogenation, CH$_2$O as the final product will be released due to its weak physical adsorption on Ni$_3$HHB, as shown in Figure 4c. The first hydrogenation of CO$_2$ is the rate-limiting step and its limiting potential is -0.62 V. The most favorable reaction path is CO$_2$ → *HCOO → *HCOOH → *CHO → CH$_2$O. Our results show that the final products on Co$_3$HIB, Cr$_3$HTB, Ti$_3$(HITP)$_2$ and V$_3$(HITP)$_2$ are CH$_4$ by eight hydrogenation steps, but their most favorable reaction pathway and rate-limiting step were different. The formation of CHOH by the CHO hydrogenation on Co$_3$HIB is rate-limiting step and its limiting potential is -0.62 V. The most favorable reaction pathway for CO$_2$RR on Co$_3$HIB is CO$_2$ → *COOH → *CO → *CHO → *CHOH → *CH$_2$OH → *CH$_3$OH → *CH$_3$ → CH$_4$. The formation of *HCOOH by the *HCOO hydrogenation on Cr$_3$HTB, Ti$_3$HITP$_2$ and V$_3$HITP$_2$ is rate-limiting step and their limiting potential are -0.29V, -0.69V, and -0.44 V, respectively. The most favorable reaction pathway for CO$_2$RR on Cr$_3$HTB, Ti$_3$HITP$_2$ and V$_3$HITP$_2$ is CO$_2$ → *HCOO → *HCOOH → *CHO → *CH2O → *CH3O → *O+CH4 → *OH → H$_2$O. In particular, a strong adsorption of H$_2$O on Ti$_3$HITP$_2$ and V$_3$HITP$_2$ results in the deactivation of TM-X$_4$ active centers, thus reducing the catalytic efficiency of these 2D MOFs.

In consequence, the C$_1$ products on five 2D MOFs and their limiting potentials (U$_L$) of their rate-limiting steps are summarized in Figure 5a. Ni$_3$HHB has the highest selectivity of the CH$_2$O product with a U$_L$ of -0.62V. Co$_3$HIB has excellent catalytic activity and selectivity with a small U$_L$ of -0.24V. Cu$_3$HIB$_2$ has the best performance for the production of HCOOH with the lowest U$_L$ of -0.03 V.

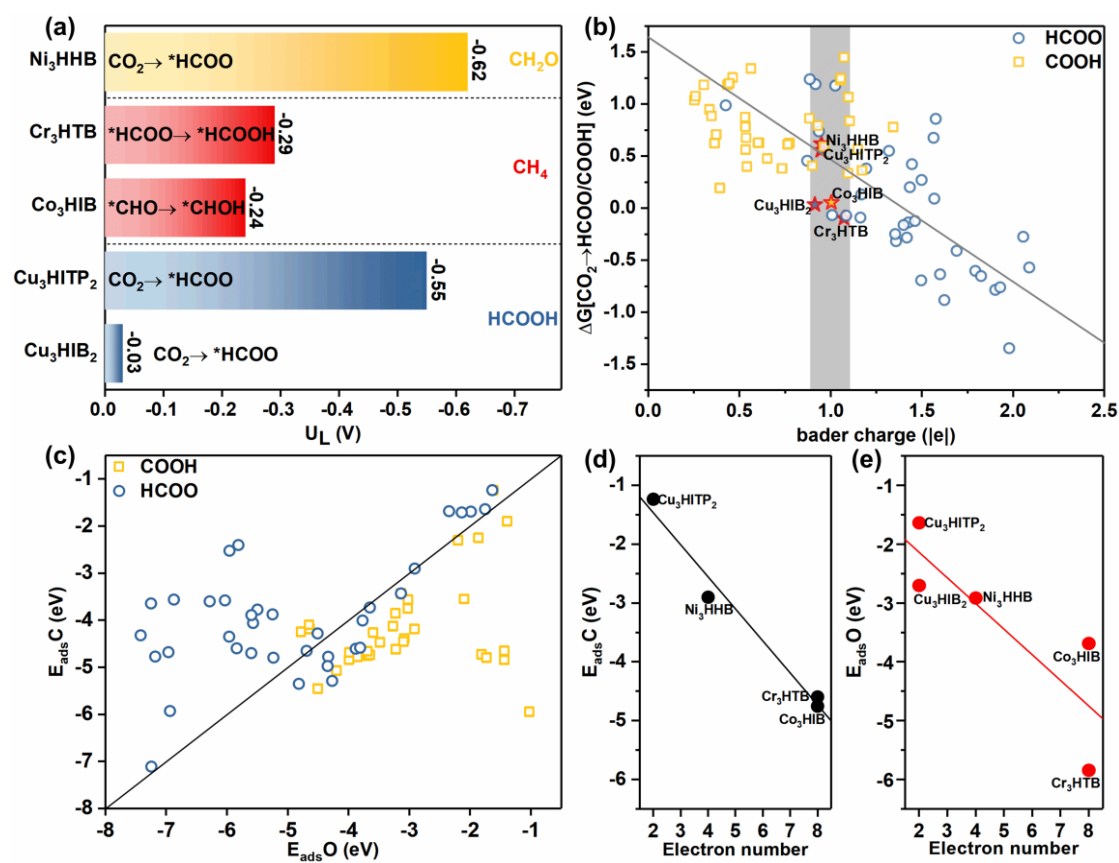

**Figure 5** (a) The limiting potentials ($U_L$) of $CO_2RR$ products and rate-limiting steps on selected five 2D MOFs as electrocatalyst candidates. (b) The linear relationship between the Bader charge of 2D MOFs and the change of Gibbs free energy of the first hydrogenation of $CO_2$ into COOH/HCOO. (c) Relationship between the adsorption energies of O and C on 2D MOFs and the selectivity of the first hydrogenation of $CO_2$ into COOH/HCOO. Linear relationship between the number of electrons involved in the $CO_2$ reduction reaction and the (d) C or (e) O adsorption energy on the catalyst surface.

**2.3 Catalytic activity and selectivity of $CO_2RR$**

The charge transfer occurs between the valence electrons of the TMs active center and the reactants or adsorbed intermediates play an important role in elementary reaction. In order to study the origin of the catalytic performance of 2D MOFs for $CO_2RR$, the relationship between the Bader charge of TMs and the change of Gibbs free energy of the first hydrogenation of $CO_2$ into COOH/HCOO was calculated, as shown in Figure 5b. The positive charge of the transition metal is regulated by the metal atom and the coordination environment. With the increase of the positive charge of the metal atom, the energy required for the $CO_2RR$ protonation process decreases and the catalytic performance is improved. However, it is worth noting that the more positive charges carried by the metal will also lead to stronger adsorption with the product and increase the difficulty of product release. In summary, too large or too small charge of metal atoms will reduce the catalytic efficiency. Interestingly, the five high-efficiency catalysts screened by high-throughput have similar metal charge of about -1.0e.

Although these five high-performance catalysts have similar metal charge, the selectivity of their $CO_2RR$ products is quite different. In the first step of $CO_2RR$ protonation, the more positively charged the active metal is the more favorable it is for $CO_2$ protonation to form *HCOO and vice versa for $CO_2$ protonation to form *COOH as shown in Figure 5b. The key factor for the formation of different intermediates is the difference in the adsorption of O and C by the catalyst; when the adsorption strength of O on the catalyst surface is much stronger than that of C, the active centers preferentially bind to O atoms and vice versa, as shown in Figure 5.5c. In addition, the stronger adsorption of the catalyst with O or C atoms will also enhance the strength of adsorption on the reaction intermediates, pushing the $CO_2RR$ towards a multi-electron reaction, as shown in Figure 5d-e.

In order to further understand the origin of catalytic activity, we studied the partial density of states of the five 2D MOFs candidate catalysts, as shown in Figure S2. In all of these catalysts, the partial-wave DOS crosses the Fermi energy level (0 eV), thus all of these candidate catalysts have metal properties and are capable of transporting electrons faster in the electrocatalytic process. In addition, the d orbitals of metals near the Fermi energy level show higher peaks compared to the p orbitals of other coordination atoms and carbon atoms, suggesting that metal atoms are more reactive. As a result, the role of the five d orbitals of the metal center in the first step of $CO_2$ protonation was comparatively analyzed, as shown in Figure 6a-e. The results show that the d orbitals of the metal atoms are acted upon by the crystal field of the coordination environment to form different splitting energy levels. Among them, $Co_3HIB$, $Cr_3HTB$ and $Ni_3HHB$ split metal d orbitals with strong hybridization near the Fermi energy level. Among them, the vertical orbital $dz^2$ is the highest energy orbital. Whereas the Cu-d orbitals of $Cu_3HITP_2$ and $Cu_3HIB_2$ split into discrete energy levels compared to $Cu_3HIB$, this change mainly depends on the long-range interactions of the carbon materials. Comparison of the changes in the metal d orbitals before and after COOH/HCOO adsorption reveals that all d orbitals are involved to some extent in the catalytic process, but the decisive role is played by the vertical orbital $dz^2$. We believe that this is due to the spatial structure of the $dz^2$ orbitals which is vertically oriented with a large number of wavefunctions concentrated at the two ends, which is more conducive to the adsorption and activation of the reaction intermediates.

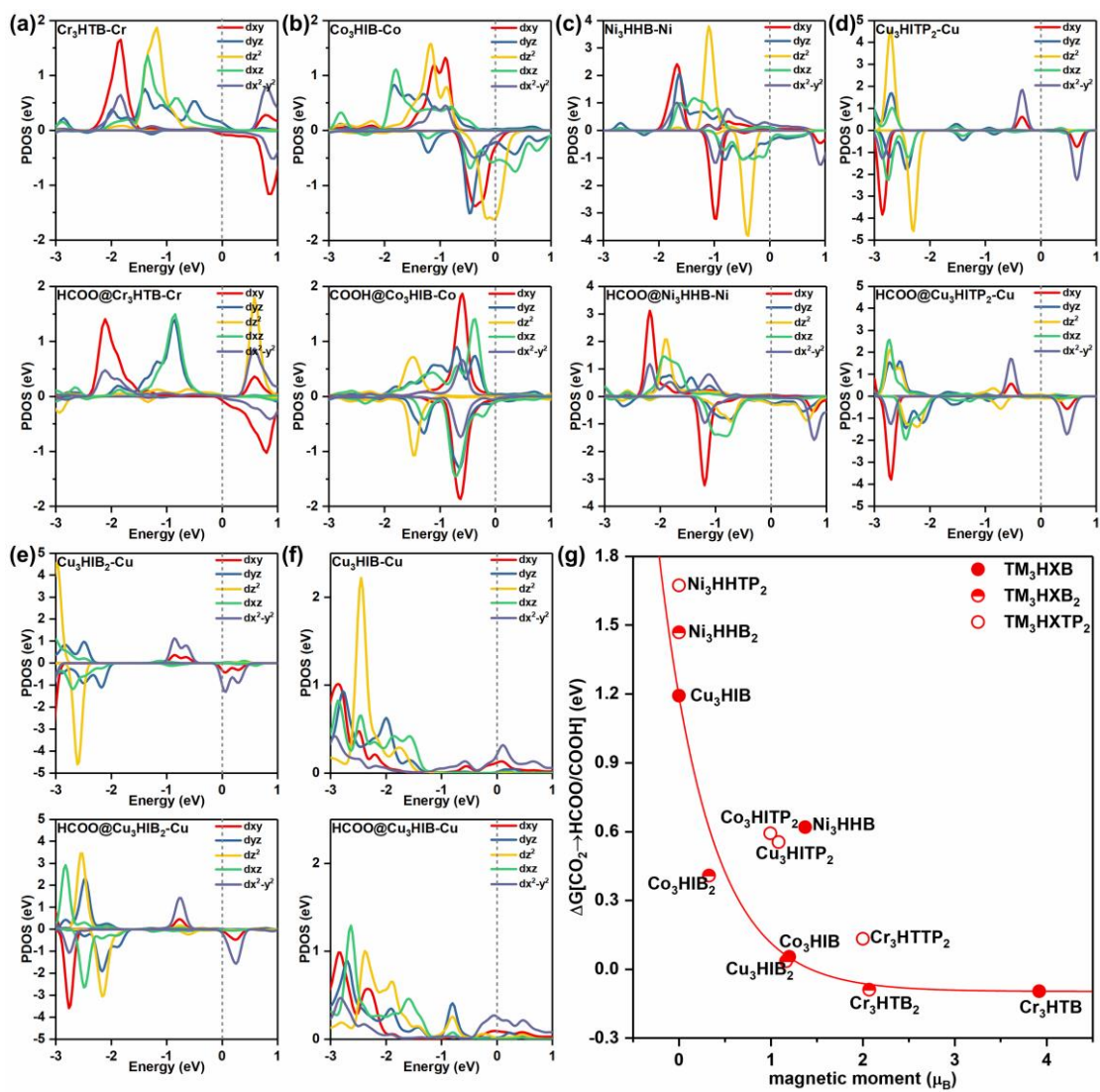

**Figure 6.** The partial density of states of metal d orbitals before and after the first step of protonation of CO$_2$RR on the (a) Cr$_3$HTB, (b)Co$_3$HIB, (c)Ni$_3$HHB, (d)Cu$_3$HITP$_2$, (e)Cu$_3$HIB$_2$, (f)Cu$_3$HIB. (g) The fitting relationship between the magnetic moment of metal atoms and the change of Gibbs free energy of the first hydrogenation of CO$_2$ into COOH/HCOO.

The interaction of different metals and organic linkers can effectively modulate the electronic structure of metal atoms and thus affect the adsorption behavior of reaction intermediates at the active center. In order to understand the reason for having the same active center but different catalytic activities, the study compared the PDOS of Cu$_3$HIB, Cu$_3$HIB$_2$, and Cu$_3$HITP$_2$ with the same Cu-N$_4$ catalytic active center as shown in Figure 6d-f. The first step protonation of all three is more inclined to produce HCOO, but the energy required for the reaction varies widely 1.20 eV, 0.03 eV, and 0.55 eV, respectively. It can be seen from the PDOS results that Cu$_3$HITP$_2$ and Cu$_3$HIB$_2$ have a strong spin polarization, and the 3d orbitals of metal

Cu cleave at the Fermi energy level to form a portion of the spin-up occupied and a partially spin-down unoccupied orbital. The ability to activate $CO_2$ is attributed to the synergistic interaction between the unoccupied and occupied d-orbitals in the metal, resulting in an "accept and donate" mechanism between the metal and the reactants.[36] In the "accept" process, electrons from the reactants are transferred to the empty d orbitals of the metal. Subsequently, the electrons in the occupied d-orbitals of the metal atom return to the antibonding orbitals of the reactants. Thus, the metal atom in the "accept and donate" process can act as an active site to weaken the C-O bond of the $CO_2$ molecule and strengthen the bonding process between the metal and the reaction intermediate. The incompletely occupied orbitals near the Fermi energy level of $Cu_3HITP_2$ and $Cu_3HIB_2$ are more conducive to the acceptance and transfer of electrons, thus improving the catalytic activity. It can also be seen that the peak at the Fermi energy level on $Cu_3HIB_2$ is shifted to the left compared to $Cu_3HITP_2$ resulting in partial occupation of the spin-down unoccupied orbitals, indicating that the Cu atoms of $Cu_3HIB_2$ are more prone to provide electrons to the reaction intermediates to promote the catalytic reaction.

Comparing the five high-performance catalysts screened above, it is found that they all have strong spin polarization phenomenon, and in order to explore the effect of spin polarization on the catalytic performance, the relationship between the catalytic activity and the spin magnetic moment of these five different pores with the same active center was further comparatively analyzed as shown in Figure 6g. The results show that the catalytic activity is lower for catalysts with zero magnetic moment of transition metal atoms in the active center. The catalytic activity increases exponentially with the increase of the spin magnetic moment of the metal atom. In previous studies, it has been demonstrated that spin state changes have a significant effect on adsorption and catalytic properties.[37, 38] However, spin modulation is usually realized by heteroatom doping and surface modification.[38, 39] Our study found that changes in pore and carbon material size also significantly affect the spin state of 2D MOFs, which will suggest new ideas for spin state modulation.

**3 Conclude**

In summary, we explored the $CO_2RR$ catalytic activity and selectivity on three types of 2D MOF materials including $TM_3HXB$, $TM_3HXB_2$, and $TM_3HXTP_2$ (TM=Ti, V, Cr, Mn, Fe, Co,

Ni, Cu; X=I, H, T, S) with different pores of TM-X$_4$ active center. By tuning the composition of the metal and ligand molecules, five high-performance CO$_2$RR catalysts with different selectivities were found, Cu$_3$HIB$_2$ and Cu$_3$HITP$_2$ with HCOOH products, Ni$_3$HHB with CH$_2$O products, and Co$_3$HIB and Cr$_3$HTB with CH$_4$ products. Synergistic interactions between the metal and ligand atoms can efficiently tune the electronic structure of the active metal. In addition, the diversity of carbon structures can modulate the spin states of metal atoms, which can affect not only the magnetic moments but also the catalytic properties of the materials. Differences in O or C affinity on the catalyst surface determine the CO$_2$RR pathway and product selectivity. Our insights will be useful for the development and optimization of highly active CO$_2$RR electrocatalysts.

**4 calculation method**

All spin-polarized density functional theory (DFT) calculations were performed by using the Vienna Ab initio Simulation Package (VASP) code.[40] The Perdew-Burke-Ernzerhof (PBE) within generalized gradient approximation (GGA) was adopted for the exchange-correlation functional and the plane-wave cutoff energy of 500 eV was employed.[41-43] The projector augmented wave (PAW) method was used to model the ion-electron interaction.[44] The convergence criterion for the energy and force were set to 10$^{-5}$ eV and 0.02 eV/Å, respectively. The DFT-D3 correction was adopted to illustrate the dispersion correction for the van der Waals (vdW) weak interactions.[45] To avoid the interactions between two adjacent periodic units, the vacuum thickness of 15 Å in the z-direction was inserted. For TM$_3$HXB, TM$_3$HXB$_2$, and TM$_3$HXTP$_2$, the Brillouin zone of geometric optimizations was sampled by a Γ-point grid of 3×5×1, 3×3×1, and 2×2×1 and the Brillouin zone of electronic properties calculations was sampled by a Γ-point grid of 7×9×1, 7×7×1, and 5×5×1, respectively.[46] An implicit solvent model is used to the effects of the water solvent environment using VASPsol with the dielectric constant set to 78.54 F/m.[47] Calculation details for the formation energy ($E_f$), the dissolution potential ($U_{diss}$), adsorption energies, Gibbs free energy, and limiting potential can be found in Supporting Information.

**Conflicts of interest**

There are no conflicts to declare.

**Acknowledgments**

The authors acknowledge the helpful discussions with Dr. Shuo Li. This study was funded by the Natural Science Foundation of China (No. 21603109), the Henan Joint Fund of the National Natural Science Foundation of China (No. U1404216), the Scientific Research Program Funded by Shaanxi Provincial Education Department (No. 20JK0676). We gratefully acknowledge the National Supercomputing Center in Zhengzhou for providing computation facilities.

**References**

[1] Xu, D.; Li, K.; Jia, B.; Sun, W.; Zhang, W.; Liu, X.; Ma, T. Electrocatalytic $CO_2$ Reduction towards Industrial Applications. *Carbon Energy* **2023**, 5 (1), e230.

[2] Zhu, Q.; Murphy, C. J.; Baker, L. R. Opportunities for Electrocatalytic $CO_2$ Reduction Enabled by Surface Ligands. *J. Am. Chem. Soc.* **2022**, 144 (7), 2829-2840.

[3] Chang, B.; Min, Z.; Liu, N.; Wang, N.; Fan, M.; Fan, J.; Wang, J. Electrocatalytic CO2 reduction to syngas. *Green Energy Environ.* **2023**.

[4] Kong, Q.; An, X.; Liu, Q.; Xie, L.; Zhang, J.; Li, Q.; Yao, W.; Yu, A.; Jiao, Y.; Sun, C. Copper-based catalysts for the electrochemical reduction of carbon dioxide: progress and future prospects. *Materials Horizons* **2023**, 10 (3), 698-721.

[5] Zhao, Z.; Lu, G. Circumventing the Scaling Relationship on Bimetallic Monolayer Electrocatalysts for Selective $CO_2$ Reduction. *Chem. Sci.* **2022**, 13 (13), 3880-3887.

[6] Zhao, C.; Su, X.; Wang, S.; Tian, Y.; Yan, L.; Su, Z. Single-atom catalysts on supported silicomolybdic acid for $CO_2$ electroreduction: a DFT prediction. *J. Mater. Chem.A* **2022**, 10 (11), 6178-6186.

[7] Wu, Q.; Si, D. H.; Liang, J.; Huang, Y. B.; Cao, R. Highly efficient electrocatalytic CO2 reduction over pyrolysis–free conjugated metallophthalocyanine networks in full pH range. *Appl. Catal. B* **2023**, 333.

[8] Gawande, M. B.; Fornasiero, P.; Zbořil, R. Carbon-Based Single-Atom Catalysts for Advanced Applications. *ACS Catal.* **2020**, 10 (3), 2231-2259.

[9] Wang, J.; Zheng, M.; Zhao, X.; Fan, W. Structure-Performance Descriptors and the Role of the Axial Oxygen Atom on M–N4–C Single-Atom Catalysts for Electrochemical CO2 Reduction. *ACS Catal.* **2022**, 12 (9), 5441-5454.

[10] Qu, G.; Wei, K.; Pan, K.; Qin, J.; Lv, J.; Li, J.; Ning, P. Emerging materials for electrochemical CO2 reduction: progress and optimization strategies of carbon-based single-atom catalysts. *Nanoscale* **2023**, 15 (8), 3666-3692.

[11] Lee, S. M.; Cheon, W. S.; Lee, M. G.; Jang, H. W. Coordination Environment in Single‐Atom Catalysts for High‐Performance Electrocatalytic $CO_2$ Reduction. *Small Structures* **2022**, 2200236.

[12] Wang, H.; Tong, Y.; Chen, P. Microenvironment regulation strategies of single-atom catalysts for advanced electrocatalytic CO2 reduction to CO. *Nano Energy* **2023**, 118.

[13] He, Q.; Lee, J. H.; Liu, D.; Liu, Y.; Lin, Z.; Xie, Z.; Hwang, S.; Kattel, S.; Song, L.; Chen, J. G. Accelerating $CO_2$ Electroreduction to CO Over Pd Single‐Atom Catalyst. *Adv. Funct. Mater.* **2020**, 30 (17), 2000407.

[14] Yang, T.; Mao, X.; Zhang, Y.; Wu, X.; Wang, L.; Chu, M.; Pao, C. W.; Yang, S.; Xu, Y.; Huang, X. Coordination tailoring of Cu single sites on $C_3N_4$ realizes selective $CO_2$ hydrogenation at low temperature. *Nat. Commun.* **2021**, 12 (1), 6022.

[15] Sun, H.; Liu, J. Carbon-supported $CoS_4$-C single-atom nanozyme for dramatic improvement in $CO_2$ electroreduction to HCOOH: A DFT study combined with hybrid solvation model. *Chin. Chem. Lett.* **2023**, 34 (8), 108018.

[16] Khan, U.; Nairan, A.; Gao, J.; Zhang, Q. Current Progress in 2D Metal–Organic Frameworks for Electrocatalysis. *Small Structures* **2022**.

[17] Zhong, H.; Wang, M.; Chen, G.; Dong, R.; Feng, X. Two-Dimensional Conjugated Metal-Organic Frameworks for Electrocatalysis: Opportunities and Challenges. *ACS Nano* **2022**, 16 (2), 1759-1780.

[18] Yang, D.; Wang, X. 2D π‐conjugated metal-organic frameworks for $CO_2$ electroreduction. *SmartMat* **2022**, 3 (1), 54-67.

[19] Wang, C.; Lv, Z.; Yang, W.; Feng, X.; Wang, B. A rational design of functional porous frameworks for electrocatalytic $CO_2$ reduction reaction. *Chem. Soc. Rev.* **2023**, 52 (4), 1382-1427.

[20] Liu, Y.; Li, S.; Dai, L.; Li, J. N.; Lv, J. N.; Zhu, Z. J. J.; Yin, A. X.; Li, P. F.; Wang, B. The Synthesis of Hexaazatrinaphthylene-Based 2D Conjugated Copper Metal-Organic Framework for Highly Selective and Stable Electroreduction of $CO_2$ to Methane. *Angewandte Chemie International Edition* **2021**, 60 (30), 16409-16415.

[21] Majidi, L.; Ahmadiparidari, A.; Shan, N.; Misal, S. N.; Kumar, K.; Huang, Z.; Rastegar, S.; Hemmat, Z.; Zou, X.; Zapol, P., et al. 2D Copper Tetrahydroxyquinone Conductive Metal-Organic Framework for Selective $CO_2$ Electrocatalysis at Low Overpotentials. *Adv. Mater.* **2021**, 33 (10), e2004393.

[22] Zhong, H.; Asl, G. M.; Ly, K. H.; Zhang, J.; Ge, J.; Wang, M.; Liao, Z.; Makarov, D.; Zschech, E.; Brunner, E., et al. Synergistic electroreduction of carbon dioxide to carbon monoxide on bimetallic layered conjugated metal-organic frameworks. *Nat. Commun.* **2020**, 11 (1), 1409.

[23] Toyoda, R.; Fukui, N.; Tjhe, D. H. L.; Selezneva, E.; Maeda, H.; Bourgès, C.; Tan, C. M.; Takada, K.; Sun, Y.; Jacobs, I., et al. Heterometallic Benzenehexathiolato Coordination Nanosheets: Periodic Structure Improves Crystallinity and Electrical Conductivity. *Adv. Mater.* **2022**, 34 (13), 2106204.

[24] Wang, Z.; Wang, G.; Qi, H.; Wang, M.; Wang, M.; Park, S.; Wang, H.; Yu, M.; Kaiser, U.; Fery, A., et al. Ultrathin Two-dimensional Conjugated Metal-organic Framework Single-crystalline Nanosheets Enabled by Surfactant-assisted Synthesis. *Chem. Sci.* **2020**, 11 (29), 7665-7671.

[25] Wang, M.; Dong, R.; Feng, X. Two-dimensional Conjugated Metal-organic Frameworks (2D c-MOFs): Chemistry and Function for MOFtronics. *Chem. Soc. Rev.* **2021**, 50 (4), 2764-2793.

[26] Liu, J.; Chen, Y.; Feng, X.; Dong, R. Conductive 2D Conjugated Metal-Organic Framework Thin Films: Synthesis and Functions for (Opto‐)electronics. *Small Structures* **2022**, 2100210.

[27] Zhong, W.; Zhang, T.; Chen, D.; Su, N.; Miao, G.; Guo, J.; Chen, L.; Wang, Z.; Wang, W. Synthesizing Cr‐Based Two‐Dimensional Conjugated Metal‐Organic Framework Through On‐Surface Substitution Reaction. *Small* **2023**, 19 (21).

[28] Hua, M.; Xia, B.; Wang, M.; Li, E.; Liu, J.; Wu, T.; Wang, Y.; Li, R.; Ding, H.; Hu, J., et al. Highly Degenerate Ground States in a Frustrated Antiferromagnetic Kagome Lattice in a Two-Dimensional Metal-Organic Framework. *J. Phys. Chem. Lett.* **2021**, 12 (15), 3733-3739.

[29] Mähringer, A.; Jakowetz, A. C.; Rotter, J. M.; Bohn, B. J.; Stolarczyk, J. K.; Feldmann, J.; Bein, T.;

Medina, D. D. Oriented Thin Films of Electroactive Triphenylene Catecholate-Based Two-Dimensional Metal‑Organic Frameworks. *ACS Nano* **2019**, 13 (6), 6711-6719.

[30] Quan, W.; Lin, Y.; Luo, Y.; Huang, Y. Y. Electrochemical $CO_2$ Reduction on Cu: Synthesis-Controlled Structure Preference and Selectivity. *Advanced Science* **2021**, 8 (23), e2101597.

[31] Wei, X.; Cao, S.; Wei, S.; Liu, S.; Wang, Z.; Dai, F.; Lu, X. Theoretical Investigation on Electrocatalytic Reduction of $CO_2$ to Methanol and Methane by Bimetallic Atoms $TM_1/TM_2$-N@Gra (TM=Fe, Co, Ni, Cu). *Appl. Surf. Sci.* **2022**, 593, 153377.

[32] Yang, L.; Pawar, A. U.; Sivasankaran, R. P.; Lee, D.; Ye, J.; Xiong, Y.; Zou, Z.; Zhou, Y.; Kang, Y. S. Intermediates and their conversion into highly selective multicarbons in photo/electrocatalytic CO2 reduction reactions. *J. Mater. Chem.A* **2023**, 11 (36), 19172-19194.

[33] Evans, M. G.; Polanyi, M. Inertia and driving force of chemical reactions. *Transactions of the Faraday Society* **1938**, 34 (0), 11-24.

[34] Bronsted, J. N. Acid and Basic Catalysis. *Chem. Rev.* **1928**, 5 (3), 231-338.

[35] Sikdar, N.; Junqueira, J. R. C.; Ohl, D.; Dieckhofer, S.; Quast, T.; Braun, M.; Aiyappa, H. B.; Seisel, S.; Andronescu, C.; Schuhmann, W. Redox Replacement of Silver on MOF-Derived Cu/C Nanoparticles on Gas Diffusion Electrodes for Electrocatalytic $CO_2$ Reduction. *Chemistry* **2022**, 28 (12), e202104249.

[36] He, C.; Wang, R.; Xiang, D.; Li, X.; Fu, L.; Jian, Z.; Huo, J.; Li, S. Charge-regulated $CO_2$ capture capacity of metal atom embedded graphyne: A first-principles study. *Appl. Surf. Sci.* **2020**, 509, 145392.

[37] Cao, A.; Nørskov, J. K. Spin Effects in Chemisorption and Catalysis. *ACS Catalysis* **2023**, 13 (6), 3456-3462.

[38] Sun, K.; Huang, Y.; Wang, Q.; Zhao, W.; Zheng, X.; Jiang, J.; Jiang, H.-L. Manipulating the Spin State of Co Sites in Metal–Organic Frameworks for Boosting CO2 Photoreduction. *Journal of the American Chemical Society* **2024**, 146 (5), 3241-3249.

[39] Gao, S.; Liu, X.; Wang, Z.; Lu, Y.; Sa, R.; Li, Q.; Sun, C.; Chen, X.; Ma, Z. Spin regulation for efficient electrocatalytic N2 reduction over diatomic Fe-Mo catalyst. *Journal of Colloid and Interface Science* **2023**, 630, 215-223.

[40] Hafner, J. Ab-initio simulations of materials using VASP: Density-functional theory and beyond. **2008**, 29 (13), 2044-2078.

[41] Perdew, J. P.; Burke, K.; Ernzerhof, M. Generalized Gradient Approximation Made Simple. *Phys. Rev. Lett.* **1996**, 77 (18), 3865-3868.

[42] Matthias, E.; Gustavo, E. S. Assessment of the Perdew-Burke-Ernzerh of Exchange-Correlation Functional. *J. Chem. Phys.* **1999**, 110 (11), 5029-5036.

[43] Kresse, G.; Furthmüller, J. Efficient iterative schemes for ab initio total-energy calculations using a plane-wave basis set. *Phys. Rev. B* **1996**, 54 (16), 11169-11186.

[44] Blöchl, P. E. Projector augmented-wave method. *Phys. Rev. B* **1994**, 50 (24), 17953-17979.

[45] Grimme, S. Semiempirical GGA-type density functional constructed with a long-range dispersion correction. **2006**, 27 (15), 1787-1799.

[46] Monkhorst, H. J.; Pack, J. D. Special points for Brillouin-zone integrations. *Phys. Rev. B* **1976**, 13 (12), 5188-5192.

[47] Kiran, M.; Ravishankar, S.; Kendra, L. W.; Arias, T. A.; Hennig, R. G. Implicit solvation model for density-functional study of nanocrystal surfaces and reaction pathways. **2014**, 140 (8), 084106.